\documentstyle[psfig]{neu}
\newcommand{\gtilde}
 {~ \raisebox{-1ex}{$\stackrel{\textstyle >}{\sim}$} ~}
\newcommand{\ltilde}
 {~ \raisebox{-1ex}{$\stackrel{\textstyle <}{\sim}$} ~}

\begin{document}

\title{%
GAMMA-RAY BURSTS FROM \\ NEUTRON STAR MERGERS AND \\ EVOLUTION OF GALAXIES}

\author{Tomonori TOTANI \\
{\it    Department of Physics, The University of Tokyo, Tokyo 113, 
JAPAN, totani@utaphp2.phys.s.u-tokyo.ac.jp}}

\maketitle

\section*{Abstract}

Most of proposed models of cosmological gamma-ray bursts (GRBs)
are associated to gravitational collapses of massive stars, and hence evolution
of the GRB rate, which is crucially important in GRB intensity distribution
analysis, is determined by the cosmic star formation history.
Here we present complementary results of GRB log$N$-log$P$ analysis, 
which were omitted
in the previous paper (Totani 1997, ApJ, 486, L71). A unique feature
of the binary neutron-star merger scenario, in contrast to
other scenarios associated to single stellar collapses, 
is that a time delay during
binary spiral-in phase emitting gravitational waves is not negligible
and makes the rate evolution flatter than that of star formation rate.
We show that the binary merger scenario is more favored than single stellar
collapses. The estimated peak luminosity and total emitted energy
in rest-frame 50-300 keV range is 1--3 $\times 10^{51} (\Omega / 4 \pi)$
erg/s and 1--3 $\times 10^{52} (\Omega / 4 \pi)$ erg, respectively,
where $\Omega$ is the opening angle of gamma-ray emission.
Absolute rate comparison between GRBs and neutron-star mergers
suggests that a beaming factor of $(\Omega/4\pi)^{-1} \sim$ a few hundreds 
is required. High-$z$ SFR data ($z>2$) based on UV luminosity
need to be corrected upwards by 
a factor of 5--10 for a good fit, and this is likely explained by the
dust extinction effect.

\section{Introduction}
The observed log$N$-log$P$
distribution of GRBs, where $N$ is the observed number of GRBs with peak
photon flux larger than $P$ [cm$^{-2}$s$^{-1}$], has been known to
agree with a cosmological distribution if the faintest bursts
are located at redshift of $z \sim 1$ and the comoving GRB rate
density is constant with time. However, cosmological evolution
of GRB occurrence rate is crucially important in the log$N$-log$P$ analysis,
and some artificial assumpsions have been made so far. Since most of 
GRB models are associated to massive stellar collapses and lifetime
of massive stars is much shorter than the cosmological time scale,
the GRB rate history is determined by the cosmic star formation history
(Totani 1997, hereafter T97; 
Sahu et al. 1997; Wijers et al. 1998). GRB models associated
to a single stellar core collapse [e.g., failed Ib supernova (Woosley 1993),
hypernova (Paczy\'{n}ski 1997)] predict GRB rate evolution simply 
proportional to star formation rate (SFR). On the other hand, rate evolution of
GRB models associated to binary systems [e.g., binary neutron-star (NS$^2$)
mergers (Blinnikov et al. 1984), accretion-induced collapse of white dwarfs
(Usov 1992)] is a little more complicated: a time delay between
star formation and GRBs makes the rate evolution flatter than that of 
SFR (T97). In the NS$^2$ merger scenario, this time delay is 
dominated by the spiral-in phase before merger emitting gravitational
waves. The duration of this phase is given as 0.0275 ($a/R_\odot)^4$ [Gyrs],
where $a$ is the initial binary separation between two neuron stars.
This strong dependence of the delay
on $a$ suggests that, a small dispersion in $a$
results in a quite wide distribution in the time delay. The time delay
becomes larger than the age of the universe when $a \gtilde 5 R_\odot$.
Considering the wide distribution of binary separation in main-sequence
binaries, this time delay is clearly not negligible in calculation of GRB rate
history. 

In the previous paper (T97), we have reported the results of 
GRB rate calculation and comparison to the 3B BATSE data, taking account
of the cosmic star formation history based on the recent observations
and the time delay in the NS$^2$ merger
scenario. Here we present some complementary results which were omitted
in T97, including results for the case that GRB rate
is simply proportional to SFR. For the details of analysis methods
or procedures, see T97.

\section{Results}

\begin{figure}[t]
  \begin{center}
    \leavevmode\psfig{figure=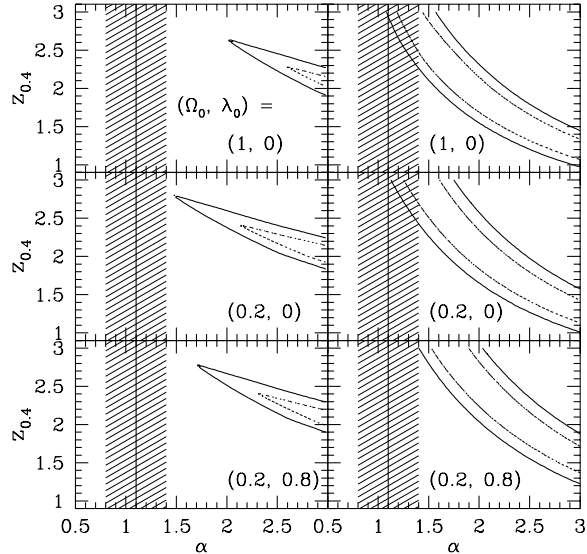,width=8cm}
  \end{center}  
   \caption{Allowed Regions for the case of GRB rate $\propto$ SFR.} 
\end{figure}

The most important fitting parameter in log$N$-log$P$ analysis is
the distance to GRBs, and here we take $z_{0.4}$: redshift of GRBs
with $P$ = 0.4 [cm$^{-2}$ sec$^{-1}$] (1024 msec). 
This peak flux is an analysis
threshold of this work. Figure 1 shows the allowed region in 
$\alpha$-$z_{0.4}$ plane with 68 \% C.L. (dotted line) and 95 \% C.L.
(solid line), for the case that GRB rate is simply proportional to
SFR, where $\alpha$ is the photon spectral index of GRBs
and shaded regions are feasible range of $\alpha$ expected from
observed GRB spectra (Mallozzi, Pendleton, \& Paciesas 1996).
Therefore the allowed region should exist in the shaded region.
This figure should be compared to Fig. 2 of T97,
in which GRBs are assumed to be NS$^2$ mergers and hence the time delay
is appropriately included.
In the left three panels, observational star formation history is assumed,
while a galaxy evolution model is used in the right panels (see T97 
for detail). The galaxy evolution model includes high-$z$ starbursts
in elliptical galaxies which have not yet been detected.
The used cosmological parameters are shown in the figure.

Compared to the NS$^2$ scenario (Fig. 2 of T97), the allowed region in Fig. 1
moves in the upper-right direction because of the steeper evolution
of SFR. Note that the comoving SFR density evolves as $(1+z)^{3.9 \pm 0.75}$ 
(Lilly et al. 1996), while NS$^2$ merger rate evolves as $(1+z)^{2-2.5}$
in $z$ = 0--1 (T97). The consequence is that the allowed region becomes
more distant from the likely range of $\alpha$ (shaded regions). We cannot
get any acceptable fit with $z_{0.4} \gtilde 2.5$ for the observational
SFR model, because of the turn over of SFR beyond $z \sim $ 2--3.
If we use the galaxy evolution model, high-$z$ starbursts may give
an acceptable fit in the shaded region with $z_{0.4} \gtilde 3$.
(We show the allowed region only in $z_{0.4} < 3$, because
it is difficult to perform a realistic comparison in $z_{0.4} \gtilde 3$
without knowledge of the epoch of elliptical galaxy formation.)
However, such high values of  $z_{0.4}$ predict higher redshift for
GRB970508. It should be noted that the estimated redshift of GRB970508
in the NS$^2$ merger scenario (T97) was already near the upper limit 
of $z$ = 2.3 (Metzger et al. 1997). Even higher redshift in the
single star scenario would be 
inconsistent with this constraint and in this case the intrinsic luminosity 
of this GRB should be significantly smaller than the average.

Cosmological time dilation analysis on GRBs gives another estimate of
GRB redshifts independent of the log$N$-log$P$ statistics. A dilation test
on the GRB duration suggested a dilation factor of $\sim$ 2.25 and redshift
of the dim bursts of $\sim 2$ (Norris et al. 1995), and this test
using GRB durations is probably the best dilation test at present
because the duration analysis is expected to be free from the
energy-dependent pulse width effect (see, e.g., Fenimore \& Bloom 1995).
This requires the GRB rate evolution of $\propto (1+z)^{1.5-2}$ 
(Horack, Emslie, \& Hartmann 1995;
Horack, Mallozzi, \& Koshut 1996; M\'{e}sz\'{a}ros \& M\'{e}sz\'{a}ros 1996),
which is well consistent with the NS$^2$ scenario.
In fact, 1024 msec peak flux of dim+dimmest bursts in Norris et al. (1995)
is 0.46 cm$^{-2}$s$^{-1}$ and our estimate of $z_{0.46}$ is $\sim$ 2--3
(see Fig. 3 of T97). The time dilation
factor for the bright and dim+dimmest bursts is
2.0--2.3 in our analysis, again in nice agreement with the result of Norris
et al. (1995).  On the other hand, in the single stellar collapse scenario,
the steeper evolution of SFR does not allow
any acceptable fit with $z_{0.4} \ltilde 3$ and even larger $z_{0.4}$ would 
make the dilation factor uncomfortably large.
Norris (1995) revised the time dilation factor from 2.25 into 1.75, and
if we believe this value, even the rate evolution in the NS$^2$ scenario 
is steep. We conclude that the modest GRB rate evolution in
the NS$^2$ merger scenario gives more natural fit to
the BATSE data than the single star scenario, although we cannot 
exclude the latter scenario completely because of
some possible uncertainties, such as the energy-dependence of the GRB
time profiles or the effect of the dispersion in the intrinsic GRB luminosity.

\section{Discussion}
We have concluded in T97 that the NS$^2$ merger scenario
gives an acceptable fit to the BATSE data, but it requires higher
SFR in $z \gtilde 2$, corresponding to high-$z$ starbursts in elliptical
galaxies. We have estimated the correction factor of 
high-$z$ SFR required to explain the missing starbursts in elliptical galaxies,
which is about a factor of 5--10. The upward correction of high-$z$ UV
flux of this degree is likey due to dust extinction. The extinction factor
is difficult to estimate, but Pettini et al. (1997) suggest a correction
factor of about 3, while Meurer et al. (1997) and Sawicki and Yee (1997)
suggest a factor of more than 10. 

Finally we estimate the peak luminosity and total emitted energy
of GRBs for the case of the NS$^2$ scenario. By using the estimated
redshifts, peak luminosity is estimated as $\sim$ 1--3 $\times 10^{51}
(\Omega/4 \pi)$ erg/sec in the rest-frame 50--300 keV, 
where $\Omega$ is the opening angle of gamma-ray
emission. Average relation between peak flux and energy fluence
of the BATSE data gives the total emitted energy of 1--3 $\times 10^{52}
(\Omega/4 \pi)$ erg
for long-duration bursts and 7 times smaller for short-duration bursts,
in the same energy range.
Absolute rate comparison between NS$^2$ mergers and the observed BATSE rate
requires $(\Omega/4 \pi)^{-1} \sim$ a few hundreds. It is very interesting
that, from an energy-budget argument of the afterglow of GRB970508,
Katz and Piran (1997) independently suggested a beaming factor of the
same order. If this beaming factor
is correct, energy required for the engine of a fireball is $\sim 10^{
50}$ erg. This energy scale as well as the beaming factor will be
useful constraints when one constructs a model of GRBs in the context of 
NS$^2$ mergers.


\section{References} \small
\re
Blinnikov, S.I., et al. 1984, Sov. Astr. Lett. 10, 177
\re
Fenimore, E.E. and Bloom, J.S. 1995, ApJ, 453, 25
\re
Horack, J. M., Emslie, A. G., \& Hartmann, D. H. 1995, ApJ, 447, 474
\re
Horack, J. M., Mallozzi, R. S., \& Koshut, T. M. 1996, ApJ, 466, 21
\re
Katz, J.I. and Piran, T. 1997, ApJ, 490, 772
\re
Lilly, S. J., F\`{e}vre, O. Le., Hammer, F., \& Crampton, D. 1996,
ApJ, 460, L1
\re
Mallozzi, R. S., Pendleton, G. N., \& Paciesas, W. S. 1996,
ApJ, 471, 636
\re
M\'{e}sz\'{a}ros, A. \& M\'{e}sz\'{a}ros, P. 1996, ApJ, 466, 29
\re
Metzger, M.R. et al. 1997, Nature, 387, 879
\re
Meurer, G.R. et al. 1997, AJ, 114, 54
\re
Norris, J.P. et al. 1995, ApJ, 439, 542
\re
Norris, J.P. 1995, in `Gamma-ray bursts: 3rd Huntsville Symposium',
eds. C. Kouveliotou, M.F. Briggs, and G.J. Fishman
\re
Paczy\'{n}ski, B. 1997, astro-ph/9706232
\re
Pettini, M. et al. 1997, astro-ph/9708117
\re
Sahu, K. et al. 1997, ApJ, 489, L127
\re
Sawicki, M \& Yee, H.K.C. 1998, AJ, in press (astro-ph/9712216)
\re
Totani, T. 1997, ApJ, 486, L71 (T97)
\re
Usov, V.V. 1992, Nature, 357, 472
\re
Wijers, R.A.M.J., et al., 1998, to appear in MNRAS (astro-ph/9708183)
\re
Woosley, S. E. 1993, ApJ, 405, 273

\end{document}